\documentstyle[12pt,epsfig]{article}

\newcounter{popnr}

\def\be{\begin{equation}}
\def\ee{\end{equation}}
\def\lsim{\raise0.3ex\hbox{$<$\kern-0.75em\raise-1.1ex\hbox{$\sim$}}}
\def\gsim{\raise0.3ex\hbox{$>$\kern-0.75em\raise-1.1ex\hbox{$\sim$}}}

\begin{document}
\title{{\bf Phenomenological studies of inclusive {\rm ep} scattering at low momentum transfer
$Q^{2}$} }

\author{
{\bf B. Surrow}\thanks{
Now at Brookhaven National Laboratory.} \thanks{Invited talk given at the International Europhysics Conference on 
High Energy Physics, July 12--18, 2001, Budapest, Hungary on behalf of the
H1 and ZEUS collaborations.} \\
Deutsches Elektronen-Synchrotron DESY \\
        Notkestr. 85, D-22607 Hamburg \\
	E-mail: bernd.surrow@desy.de}
\date{}
\maketitle
\thispagestyle{empty}

\abstract{Phenomenological studies on the proton structure function $F_{2}$
have been carried out to investigate the behavior of $F_{2}$, 
i.e. its $x$ and $Q^{2}$ dependence, in the transition from the deep-inelastic 
scattering region ($Q^{2}\gg 1\,$GeV$^{2}$) to the photoproduction region
($Q^{2} \approx 0\,$GeV$^{2}$). An overview of various results of $F_{2}$ at low 
$Q^{2}$ is given. To quantify the behavior of $F_{2}$ at low $Q^{2}$, the derivatives
$d\ln(F_{2})/d\ln(x)$ for fixed $Q^{2}$ and $dF_{2}/d\log (Q^{2})$ for fixed $x$ 
have been determined and compared to expectations within the framework of perturbative
and non-perturbative QCD. Furthermore, the derivatives were compared to results
of the total inclusive diffractive cross-section and vector-meson production. 
An interpretation of the inclusive {\rm ep} results is given in the framework 
of the GVD/CDP picture with special emphasis on the limit of $Q^{2} \rightarrow 0$
for fixed $W^{2}$ and $x \rightarrow 0$ for fixed $Q^{2}$.}

\newpage

\addtocounter{page}{-1}


\section{Introduction}

The measurement of the proton structure function $F_{2}$ at low $Q^{2}$
at HERA has attracted lots of interest and attention in recent years. 
The H1 and ZEUS Collaborations have measured $F_{2}$ at low $Q^{2}$ with high
precision using various experimental techniques. This allowed to study
the behavior of $F_{2}$ in the transition from the deep-inelastic 
scattering region ($Q^{2}\gg 1\,$GeV$^{2}$) to the photoproduction region
($Q^{2} \approx 0\,$GeV$^{2}$).

At low $Q^{2}$ a perturbative QCD (pQCD) approach based on NLO DGLAP fits
is expected to lose its validity for fundamental reason. This is due to the underlying
non-Abelian nature of the QCD field theory and its corresponding behavior of 
the coupling constant $\alpha_{s}$ to be large at large distances which therefore
does not permit a perturbative approach. Thus other non-perturbative QCD
concepts have to be considered such as Regge phenomenology, Generalized Vector
Dominance (GVD) and Color Dipole Picture (CDP) model approaches to name only the most prominent. 

It has been shown that a pQCD analysis based on NLO DGLAP fits allow to describe
the behavior of $F_{2}$ down to approximately $Q^{2} \approx 1\,$GeV$^{2}$ \cite{ref1}.
Although the quality of those fits are still acceptable
down to $Q^{2} \approx 1\,$GeV$^{2}$, various expectations on the behavior
of the underlying parton distributions at high $Q^{2}$ are no longer valid 
at low $Q^{2}$, e.g. the behavior of the gluon distribution. This gave rise to
several theoretical debates on the validity of NLO DGLAP fits at low $Q^{2}$
indicating that the limits of a pQCD approach are reached earlier than simply
given by the quality of the NLO DGLAP fits. Putting it in other words, the
application of a partonic picture such as pQCD based on NLO DGLAP fits 
cannot necessarily be pushed down to $Q^{2} \approx 1\,$GeV$^{2}$ without
loosing its physical interpretation. At low $Q^{2}$ and in particular 
in the limit of $Q^{2} \rightarrow 0$ - which is a well known fact -
a hadronic behavior becomes important which can be well accounted for
by Regge phenomenology.

It has recently been shown that the GVD/CDP model approach provides at low $x$
a successful description of $F_{2}$ from the deep-inelastic 
scattering region ($Q^{2}\gg 1\,$GeV$^{2}$) to the photoproduction region
($Q^{2} \approx 0\,$GeV$^{2}$). It still remains to be shown how these findings
can be interpreted based on a microscopic picture of quark and gluon dynamics. Such
an attempt has recently been published in \cite{ref2}. 

The next section will provide an overview of various $F_{2}$ results showing
the $Q^{2}$ and $x$ dependence of $F_{2}$ as well as the $Q^{2}$ and $W^{2}$
dependence of the total $\gamma^{*} p$ cross-section $\sigma_{tot}^{\gamma^{*}p}$
which is in particular important when approaching the photoproduction limit 
($Q^{2} \rightarrow 0$). 

To quantify the behavior of $F_{2}$ at low $Q^{2}$, the derivatives
$d\ln(F_{2})/d\ln(x)$ for fixed $Q^{2}$ and $dF_{2}/d\log (Q^{2})$ for fixed $x$ 
have been determined and compared to expectations within the framework of perturbative
and non-perturbative QCD. Furthermore, the derivatives were compared to results
of the total inclusive diffractive cross-section and vector-meson production.
Those results will be shown in section 3.  

An attempt is made in section 4 to interpret the HERA inclusive {\rm ep} scattering
results in the framework of the GVD/CDP model approach. It will be shown that
a universal hadronic behavior of the total $\gamma^{*} p$ cross-section is reached not
only in the limit of $Q^{2} \rightarrow 0$ for fixed $W^{2}$ which is seen at
HERA but also - without experimental verification yet - in the limit of
$x \rightarrow 0$ for fixed $Q^{2}$, i.e. 
$\sigma_{tot}^{\gamma^{*}p} \rightarrow \sigma_{tot}^{\gamma p}$. 

\section{Overview plots of $F_{2}$ and $\sigma_{tot}^{\gamma^{*}p}$}

\begin{figure}[t]
\centerline{\epsfig{figure=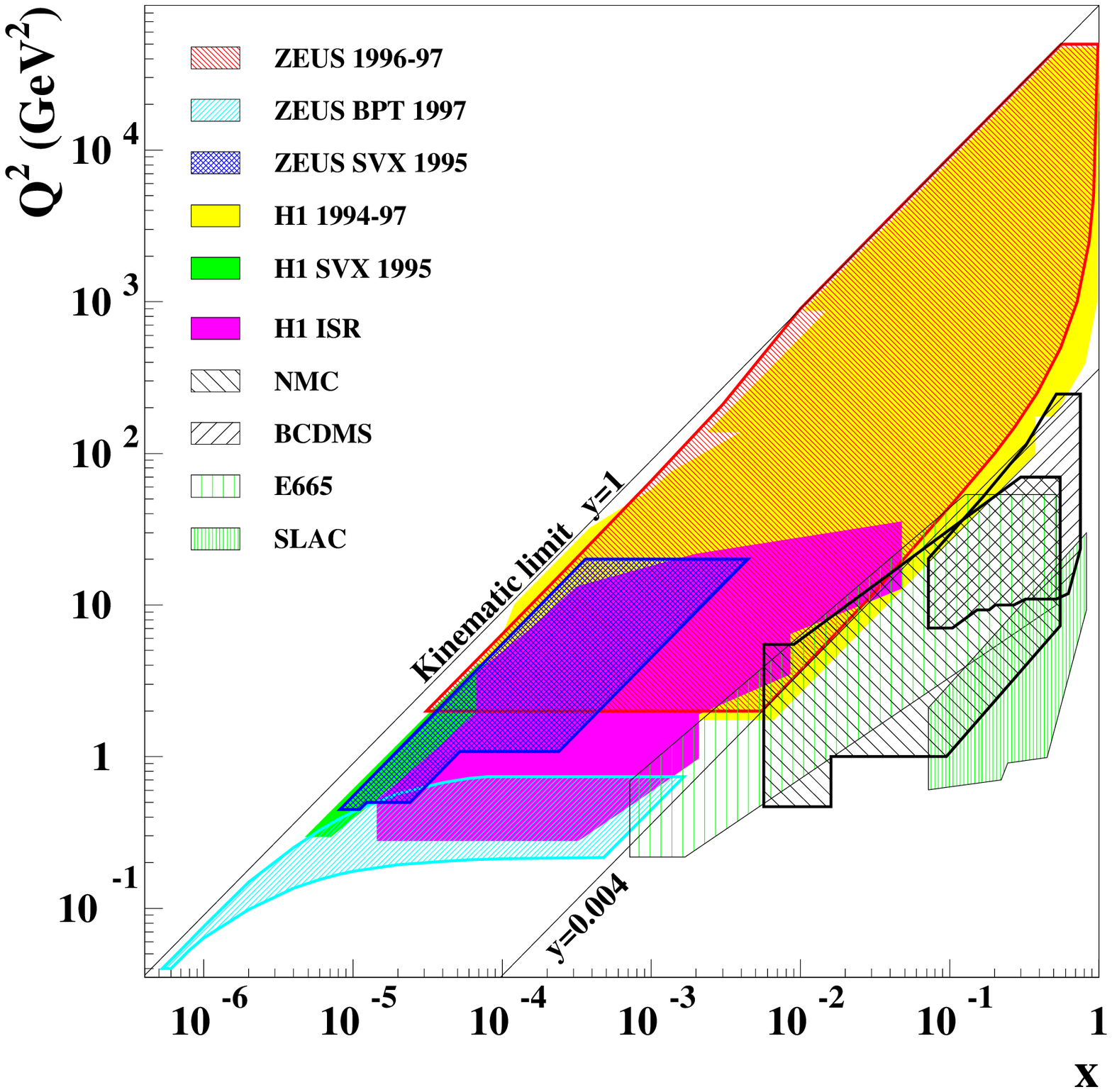,width=15.0cm}}
\caption{\it Kinematic coverage in the $Q^{2}-x$ plane for various fixed-target
experiments and the HERA collider experiments H1 and ZEUS.}
\label{kin_plane}
\end{figure}

Figure \ref{kin_plane} shows the kinematic coverage in the $Q^{2}-x$ plane for 
various fixed-target experiments and the HERA collider experiments H1 \cite{ref3a} and ZEUS \cite{ref3b}.
Access to the low $Q^{2}$ region was made possible using various experimental techniques.
A dedicated detector to measure the scattered electron under very small angles and thus
to provide acceptance at low $Q^{2}$ has been installed within the ZEUS experiment. This
effort allowed to measure $F_{2}$ over a wide range in $x$ and $Q^{2}$ with high precision 
(ZEUS BPT 1997). Dedicated HERA runs with a shifted event vertex compared to the 
nominal interaction vertex have been used at H1 and ZEUS to effectively gain acceptance
at low $Q^{2}$ (H1 SVX 1995 and ZEUS SVX 1995). Preliminary results on a measurement of
$F_{2}$ at low $Q^{2}$ using initial-state radiation events thus using events with a
lower center-of-mass energy and thereby gaining acceptance at low $Q^{2}$ has been 
reported by the H1 Collaboration (H1 ISR). 

\begin{figure}[b]
\centerline{\epsfig{figure=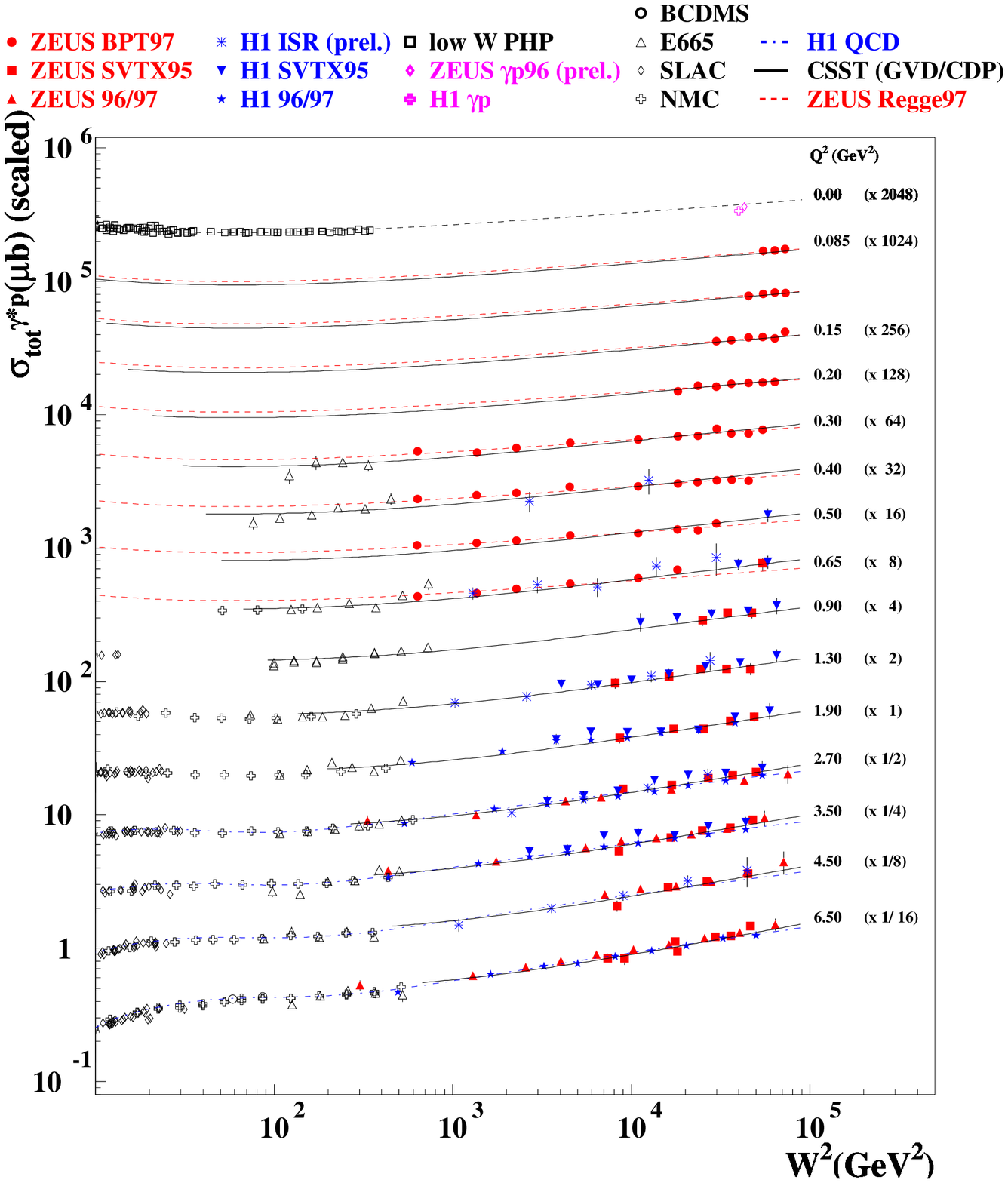,width=15.0cm}}
\caption{\it $\sigma_{tot}^{\gamma^{*} p}$ as a function of $W^{2}$ for different values
of $Q^{2}$.}
\label{sig_w2}
\end{figure}

\begin{figure}[t]
\centerline{\epsfig{figure=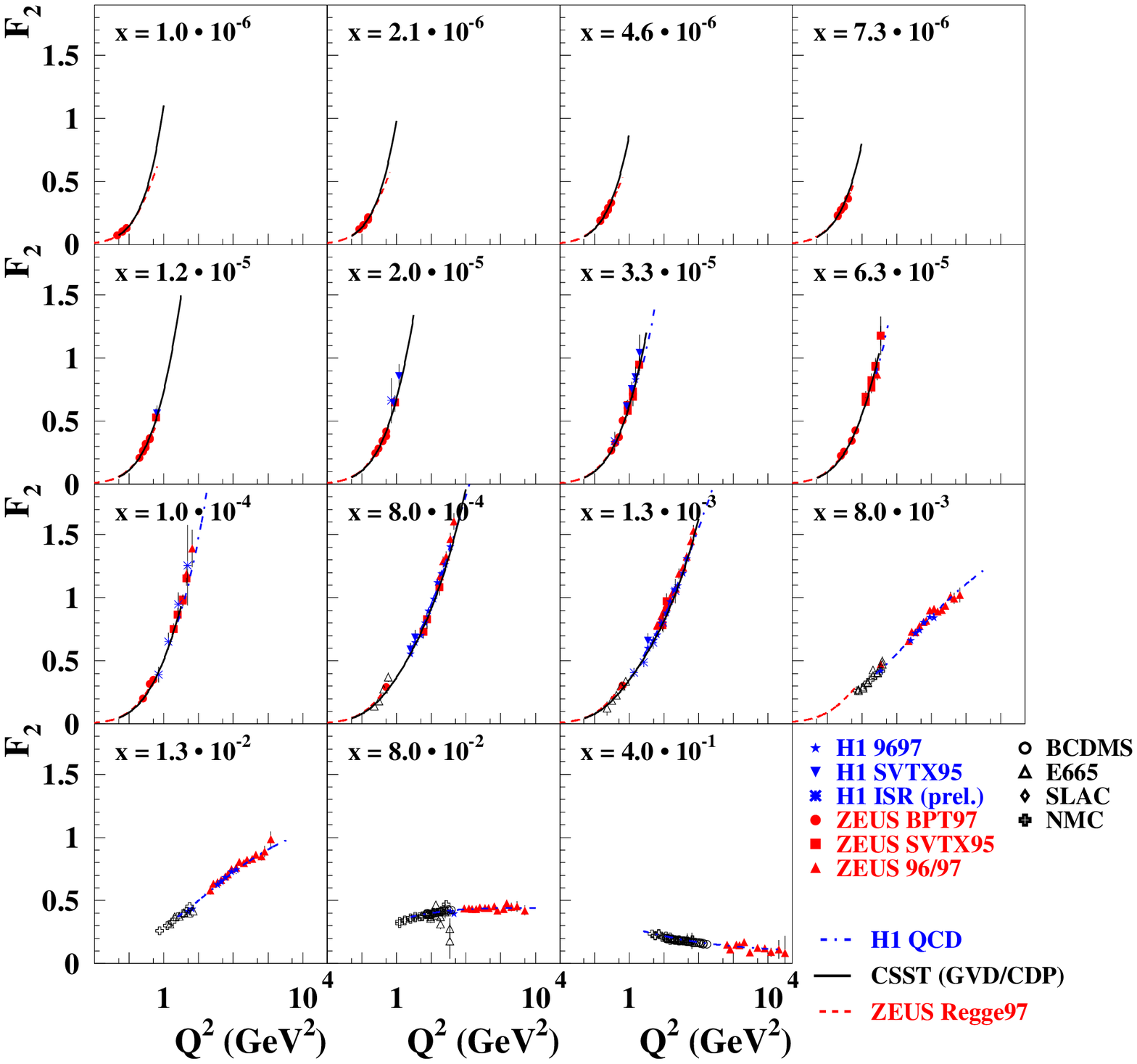,width=15.0cm,clip=}}
\caption{\it $F_{2}$ as a function of $Q^{2}$ for different values
of $x$.}
\label{f2_q2}
\end{figure}

\begin{figure}[h]
\centerline{\epsfig{figure=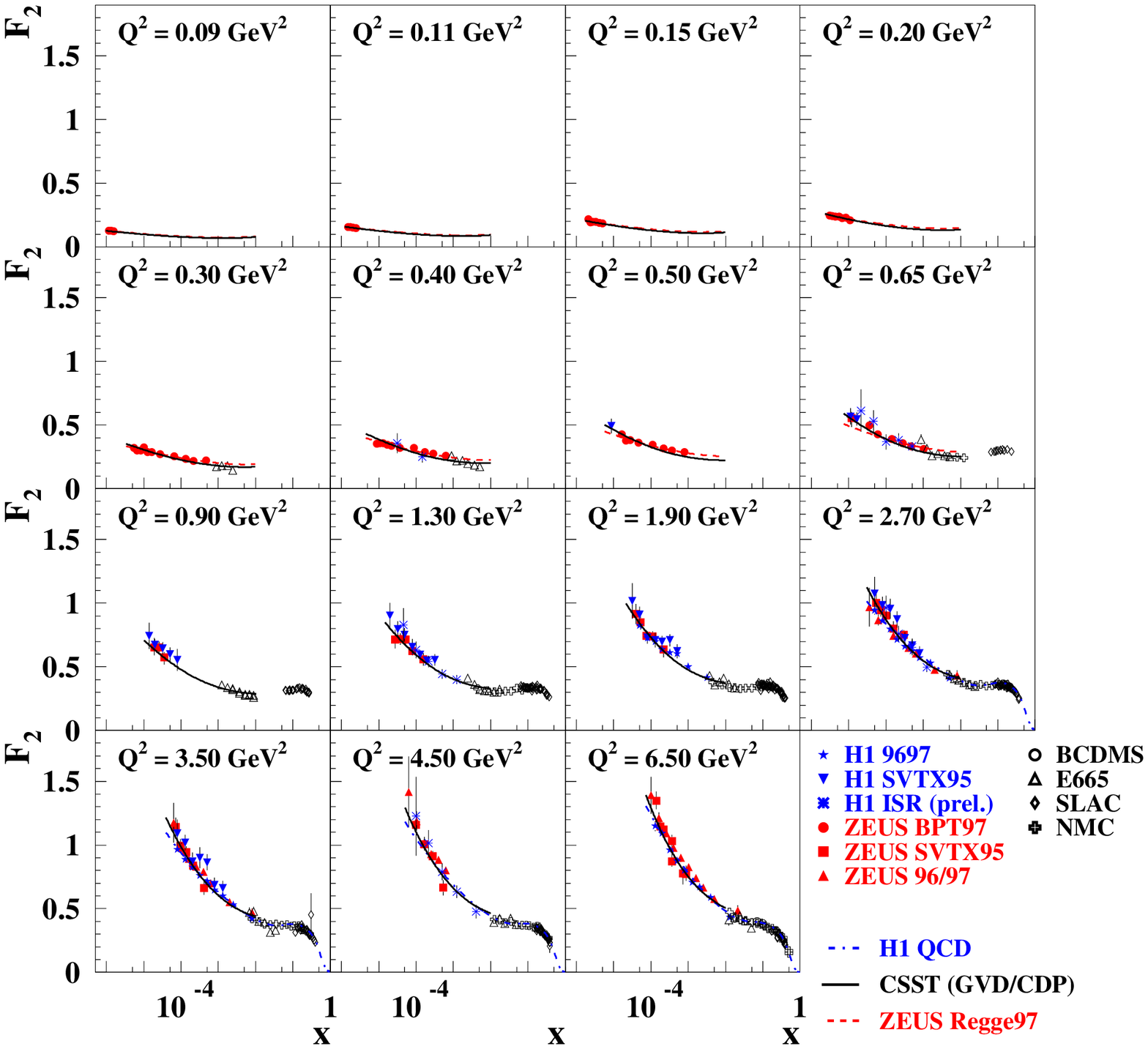,width=15.0cm}}
\caption{\it $F_{2}$ as a function of $x$ for different values
of $Q^{2}$.}
\label{f2_x}
\end{figure}

Those measurements allowed an experimental basis to examine the behavior of $F_{2}$
at low $Q^{2}$ over a wide range in $x$. Taking all these measurements together
overview plots of the $Q^{2}$ and $x$ dependence of $F_{2}$ as well as the $Q^{2}$ and $W^{2}$
dependence of the total $\gamma^{*} p$ cross-section $\sigma_{tot}^{\gamma^{*}p}$
are shown in Figures \ref{sig_q2}, \ref{sig_w2}, \ref{f2_q2} and \ref{f2_x}.

Figure \ref{sig_q2} shows the total $\gamma^{*} p$ cross-section 
as a function of $Q^{2}$ for different values of $W$. Also shown 
is the measurement of the total photoproduction cross-section by the
H1 and ZEUS Collaborations. A steady decrease of
$\sigma_{tot}^{\gamma^{*}p}$ with increasing $Q^{2}$ for $Q^{2}>1\,$GeV$^{2}$
can be seen which is well accounted for by NLO DGLAP-fits \cite{ref3a} (H1 97). Going towards
lower values of $Q^{2}$ a characteristic change of $\sigma_{tot}^{\gamma^{*}p}$
is visible at around $Q^{2}\approx 1\, $GeV$^{2}$. The region of $Q^{2}<1\,$GeV$^{2}$ 
can be well represented by a Regge/VDM-fit \cite{ref3b} (ZEUS BPT 1997). 
A GVD/CDP-model approach using the ansatz by
\cite{ref4} allows to describe $\sigma_{tot}^{\gamma^{*}p}$ for $Q^{2}<100\,$GeV$^{2}$
and $x<0.01$. 

Figure \ref{sig_w2} shows the total $\gamma^{*} p$ cross-section 
as a function of $W^{2}$ for different values of $Q^{2}$. 
$\sigma_{tot}^{\gamma^{*}p}$ is seen to increase more strongly with $W^{2}$
for larger values of $Q^{2}$ which is well described by NLO DGLAP-fits \cite{ref3a} (H1 97). At low
$Q^{2}$ and thus towards the photoproduction limit the milder increase of 
$\sigma_{tot}^{\gamma^{*}p}$ can be represented by a Regge/VDM-fit \cite{ref3b} (ZEUS BPT 1997). For $x<0.01$
and $Q^{2}<100\,$GeV$^{2}$ the GVD/CDP-model approach \cite{ref4} provides
a good description of the $W^{2}$ dependence.

Figure \ref{f2_q2} shows $F_{2}$ 
as a function of $Q^{2}$ for different values of $x$.  $F_{2}$ diminishes towards
low $Q^{2}$ like $F_{2} \propto Q^{2} \rightarrow 0$ for $Q^{2} \rightarrow 0$.
The rise of $F_{2}$ with $Q^{2}$ increases in the low $x$ region which is larger at
high $Q^{2}$ and smaller at low $Q^{2}$. The region of $Q^{2} < 1\, $GeV$^{2}$
can be well described by a Regge/VDM-fit \cite{ref3b} (ZEUS BPT 1997) whereas the region for $Q^{2} > 1\, $GeV$^{2}$
can be well accounted for by NLO-DGLAP fits \cite{ref3a} (H1 97). For $x<0.01$
and $Q^{2}<100\,$GeV$^{2}$ the GVD/CDP-model approach \cite{ref4} provides
a good description of the $Q^{2}$ dependence.

Figure \ref{f2_x} shows $F_{2}$ 
as a function of $x$ for different values of $Q^{2}$. The rise of $F_{2}$ 
diminishes towards low $Q^{2}$ and reaches the expected $x$-dependence which is
well accounted for by a Regge/VDM-description \cite{ref3b} (ZEUS BPT 1997) based on a soft Pomeron
ansatz. Towards larger values of $Q^{2}$ NLO DGLAP-fits \cite{ref3a} (H1 97) successfully describe 
the $x$-dependence of $F_{2}$. For $x<0.01$
and $Q^{2}<100\,$GeV$^{2}$ the GVD/CDP-model approach \cite{ref4} provides
a good description of the $x$ dependence.

\section{Phenomenological studies on the $x$ and $Q^{2}$ dependence}

Both, the H1 and ZEUS Collaborations have carried out phenomenological studies
to quantify the observed behavior of $F_{2}$ at low $Q^{2}$ which was
discussed in the last section. The derivatives
$d\ln(F_{2})/d\ln(x)$ for fixed $Q^{2}$ and $dF_{2}/d\log (Q^{2})$ for fixed $x$ 
have been determined and compared to expectations within the framework of perturbative
and non-perturbative QCD.

\begin{figure}[ht]
\centerline{\epsfig{figure=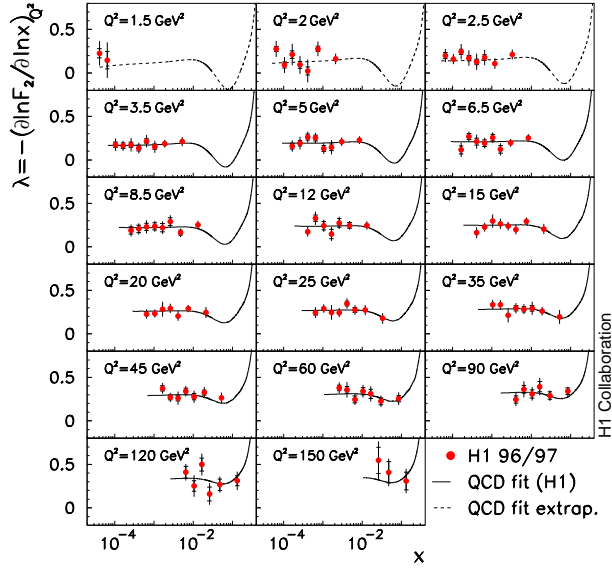,width=8.0cm}}
\caption{\it H1 results on the measurement of the function $\lambda(x,Q^2)$ as a function of $x$ for different
values of $Q^{2}$.}
\label{lambda_x_h1}
\end{figure}

The H1 Collaboration extracted for the first time the derivative 
$\left( \frac{\partial \ln F_2(x,Q^2)}{\partial \ln x}\right)_{Q^2}  \equiv 
    - \lambda(x,Q^2)$
as a function of $Q^{2}$ and $x$ \cite{ref5}. As can be seen from Figure \ref{lambda_x_h1},
the derivative $\lambda(x,Q^2)$ is found to be independent of $x$ for $x<0.01$. This 
implies that the $x$ dependence of $F_{2}$ at low $x$ is consistent with a power-law
behavior, $F_{2} \propto x^{-\lambda}$, for fixed values of $Q^{2}$. Furthermore the
rise of $F_{2}$ at low $x$ which was one of the first observations at HERA, is found
to be proportional to $F_{2}/x$. Thus, there is no evidence for a change in this
behavior in the measured kinematical region. The observed behavior at low $x$ is
consistent with the result of a pQCD fit.

\begin{figure}[ht]
\centerline{\epsfig{figure=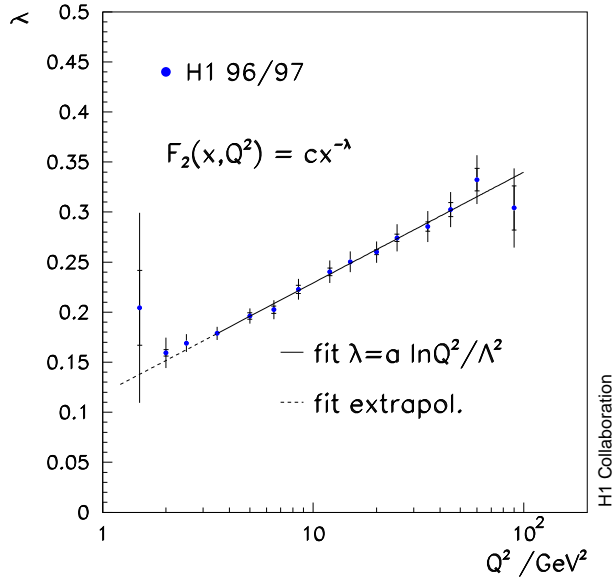,width=8.0cm}}
\caption{\it H1 results on $\lambda(Q^{2})$ as a function of $Q^{2}$ extracted from a fit
of the form $F_{2}=C(Q^{2})\cdot x^{-\lambda(Q^{2})}$.}
\label{lambda_q2_h1}
\end{figure}

These findings justify that the behavior of $F_{2}$ can be simply  
quantified as $F_{2}=C(Q^{2})\cdot x^{-\lambda(Q^{2})}$. Figure \ref{lambda_q2_h1}
shows the exponent $\lambda(Q^{2})$ as a function of $Q^{2}$. $\lambda(Q^{2})$ is seen to rise
linearly with $\log Q^{2}$. The extracted coefficient $C(Q^{2})$ is found to be independent
of $Q^{2}$ within experimental uncertainties.  

\begin{figure}[ht]
\centerline{\epsfig{figure=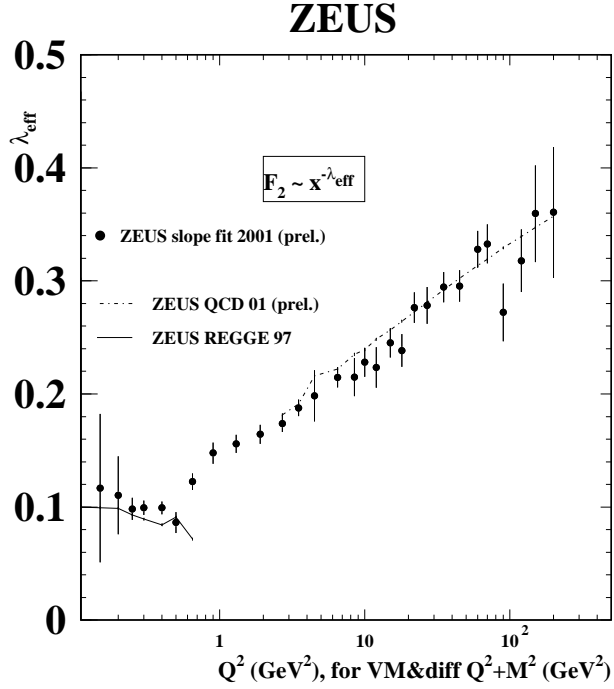,width=8.0cm}}
\caption{\it ZEUS preliminary results on $\lambda_{eff}=\lambda(Q^{2})$ as a function of $Q^{2}$. The estimates
of the ZEUS Regge97 and ZEUS QCD01 parameterizations are also shown.}
\label{lambda_q2_zeus}
\end{figure}

A similar analysis extracting $\lambda(Q^{2})$ as a function of $x$ has been carried
out by the ZEUS Collaboration. The result of this preliminary analysis is shown in Figure 
\ref{lambda_q2_zeus}. The extracted values for $\lambda(Q^{2})$ extend to lower values in 
$Q^{2}$. The results agree with the ZEUS QCD01 parameterization down to $Q^{2}=2.5\,$GeV$^{2}$.
For $Q^{2}<0.6\,$GeV$^{2}$, $\lambda(Q^{2})$ is consistent with a constant value of $\approx 0.1$
as expected from the assumption of a single soft Pomeron exchange within the framework
of Regge phenomenology. Thus the slope in $F_{2}$ at low $x$ and very low $Q^{2}$ is found to
be independent of $Q^{2}$ within experimental uncertainties. This behavior can be also inferred
from Figure \ref{sig_q2}. The $\gamma^{*}p$ cross-section at fixed values of $W^{2}$ 
saturates for $Q^{2} \rightarrow 0$.

An attempt was made to compare the energy dependence of the inclusive $\gamma^{*}p$ cross-section
which is equivalent to the above studies on the $x$ dependence of $F_{2}$, to the
energy dependence of the total inclusive diffractive cross-section (Diff) , vector meson production (VM)  
and deeply virtual Compton scattering (DVCS) at HERA.
The total $\gamma^{*}p$ cross-section can be related to the elastic vector meson cross-section
using the Optical Theorem which implies that $\sigma_{tot}^{\gamma^{*}p}$ varies with $W^{2}$ as
$\sigma_{tot}^{\gamma^{*}p} \propto (W^{2})^{\epsilon}$ whereas the elastic vector meson 
cross-section varies with $W^{2}$ as 
$\sigma_{el}(\gamma^{*} p \rightarrow Vp) \propto (W^{2})^{2\epsilon}$. 

\begin{figure}[ht]
\centerline{\epsfig{figure=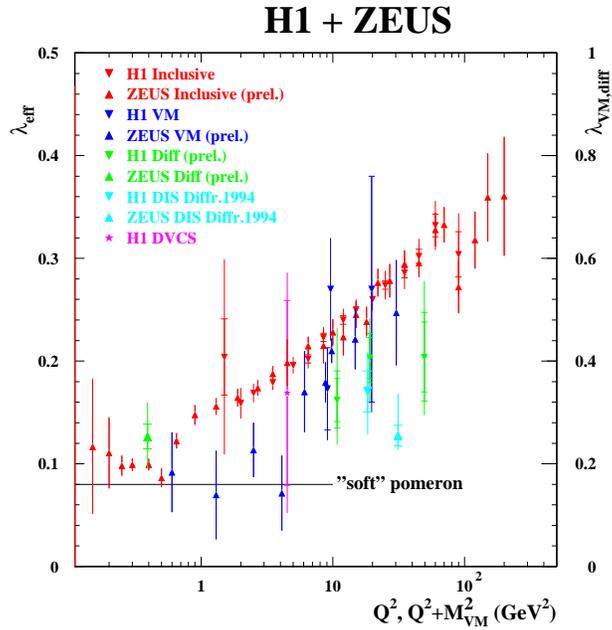,width=8.0cm}}
\caption{\it Compilation of various slope analyses from the
H1 and ZEUS Collaborations on the energy dependence of the respective cross-sections:
inclusive results (Inclusive), vector-meson results (VM), inclusive diffractive results (Diff) and 
deeply-virtual Compton scattering results (DVCS).}
\label{lambda_q2_all}
\end{figure}

Figure \ref{lambda_q2_all} shows a compilation of various slope analyses from the
H1 and ZEUS Collaborations. The inclusive results which were discussed earlier are
found to be consistent within experimental uncertainties. The vector meson analysis 
indicated that the equivalent slope, $\lambda_{VM}$, scales with $Q^{2}+M^{2}_{VM}$ where
$M_{VM}$ is the vector meson mass. The extracted values for $\lambda_{Diff}$, 
$\lambda_{VM}$ and $\lambda_{DVCS}$ are multiplied by a factor $1/2$ in the comparison
to the inclusive results. 

The naive expectations of $\lambda_{VM}/2=\lambda$ is approximately true.
$\lambda_{Diff}/2$ is somewhat below the inclusive result indicating that the
inclusive and inclusive diffractive cross-sections have the same $W$ dependence contrary
to the case of elastic vector-meson production.

The large uncertainties however do not permit to draw any further conclusions yet. 

\begin{figure}[ht]
\centerline{\epsfig{figure=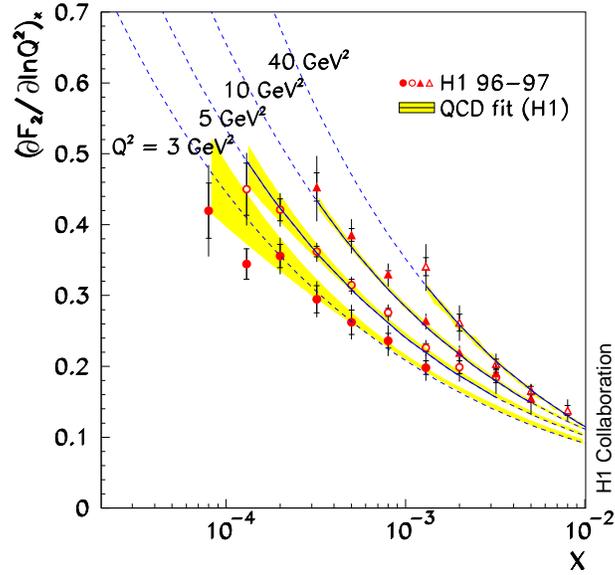,width=8.0cm}}
\caption{\it H1 results on $(\partial F_{2}/ \partial \ln Q^{2})_{x}$ as function of $x$ for different
values of $Q^{2}$. The prediction from a QCD-fit is also shown.}
\label{df2dloq2_x_h1}
\end{figure}

\begin{figure}[ht]
\centerline{\epsfig{figure=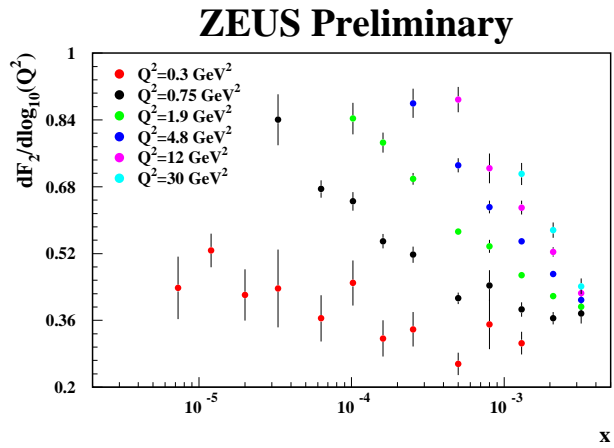,width=8.0cm}}
\caption{\it ZEUS preliminary results on $(\partial F_{2}/ \partial \log Q^{2})_{x}$ as function of $x$ for different
values of $Q^{2}$.}
\label{df2dloq2_x_zeus}
\end{figure}

Both, the H1 and ZEUS Collaborations, extracted besides the $x$ slope of $F_{2}$
also the $Q^{2}$ slope, i.e. $(\partial F_{2}/ \partial \log Q^{2})_{x}$. Those
results are shown in Figure \ref{df2dloq2_x_h1} \cite{ref3a} (H1 97) and \ref{df2dloq2_x_zeus}. A consistent
picture of the H1 and ZEUS analysis is obtained on the $Q^{2}$ slope of $F_{2}$ as a function of
$x$. At low $Q^{2}$, the slope of $F_{2}$ increases only slowly towards low values of $x$
whereas at high $Q^{2}$ it increases rather strongly. This behavior can be also inferred
from Figure \ref{f2_q2} which is in agreement with pQCD predictions as shown in Figure 
\ref{df2dloq2_x_h1}.

\section{Interpretation of HERA data}

The $W^{2}$ and $Q^{2}$ dependence of the total $\gamma^{*}p$ cross-section
has been discussed in detail in the last section. 

To summarize, the steep rise of the inclusive $\gamma^{*}p$ cross-section
with increasing $W^{2}$ which is equivalent to the steep rise of $F_{2}$ towards
low $x$, diminishes for $Q^{2} \rightarrow 0$ and reaches the energy dependence of the
total $\gamma p$ cross-section. At low $Q^{2}$, the cross-section behavior of the
total inclusive $\gamma^{*}p$ cross-section can be well accounted for by Regge phenomenology
whereas at high $Q^{2}$, pQCD based on a NLO DGLAP analysis allows to describe the
observed behavior. 

At low $x$ and $Q^{2}<100\,$GeV$^{2}$ it was shown that a GVD/CDP-model approach 
allows to describe the observed cross-section behavior. 

In the formulation of the above model by \cite{ref4}, it has been shown
that virtual as well as real photon cross-sections show a scaling behavior of the form
$\sigma_{tot}^{\gamma^{(*)}p}(W^{2},Q^{2})=\sigma_{tot}^{\gamma^{(*)}p}(\eta)$ 
with $\eta=(Q^{2}+m_{0}^{2})/\Lambda(W^{2})$. 
For related approaches compare \cite{ref6}.

\begin{figure}[ht]
\centerline{\epsfig{figure=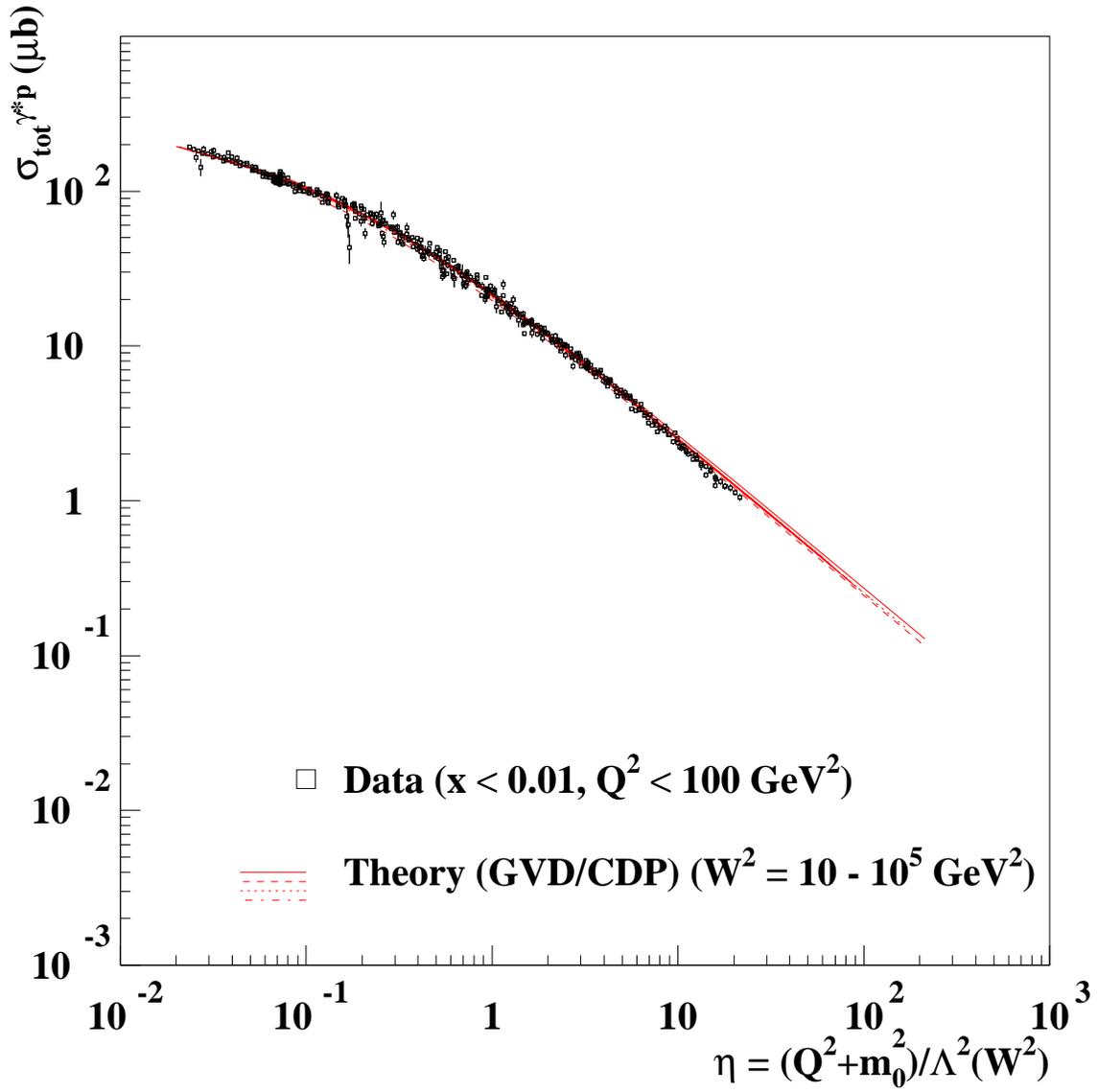,width=15.0cm}}
\caption{\it The experimental data for $\sigma_{tot}^{\gamma^{(*)}p}(W^{2},Q^{2})$
for $x<0.01$ including photoproduction results as a function of the scaling variable 
$\eta=(Q^{2}+m_{0}^{2})/\Lambda(W^{2})$.}
\label{sig_gvdm_eta}
\end{figure}

The scale $\Lambda^{2}(W^{2})$ turned out to be
an increasing function of $W^{2}$ and my be represented by a power law 
($\Lambda^{2}(W^{2})=C_{1}(W^{2}+W_{0}^{2})^C_{2}$) or logarithmic 
($\Lambda^{2}(W^{2})=C^{'}_{1}\ln(W^{2}/W_{0}^{2}+C^{'}_{2})$) 
function of $W^{2}$. The threshold mass $m_{0}^{2}$ as well as the other three parameters
of the function $\Lambda^{2}(W^{2})$ have been constrained by the data itself. The result of
this analysis can be seen in Figure \ref{sig_gvdm_eta}. It has been shown that this scaling behavior
improves with decreasing $x$. The HERA data on DIS in the low $x$ diffraction region, including
photoproduction, find a natural interpretation in the GVD/CDP picture that rests on the generic
structure of two-gluon exchange from QCD. 

\begin{figure}[t]
\centerline{\epsfig{figure=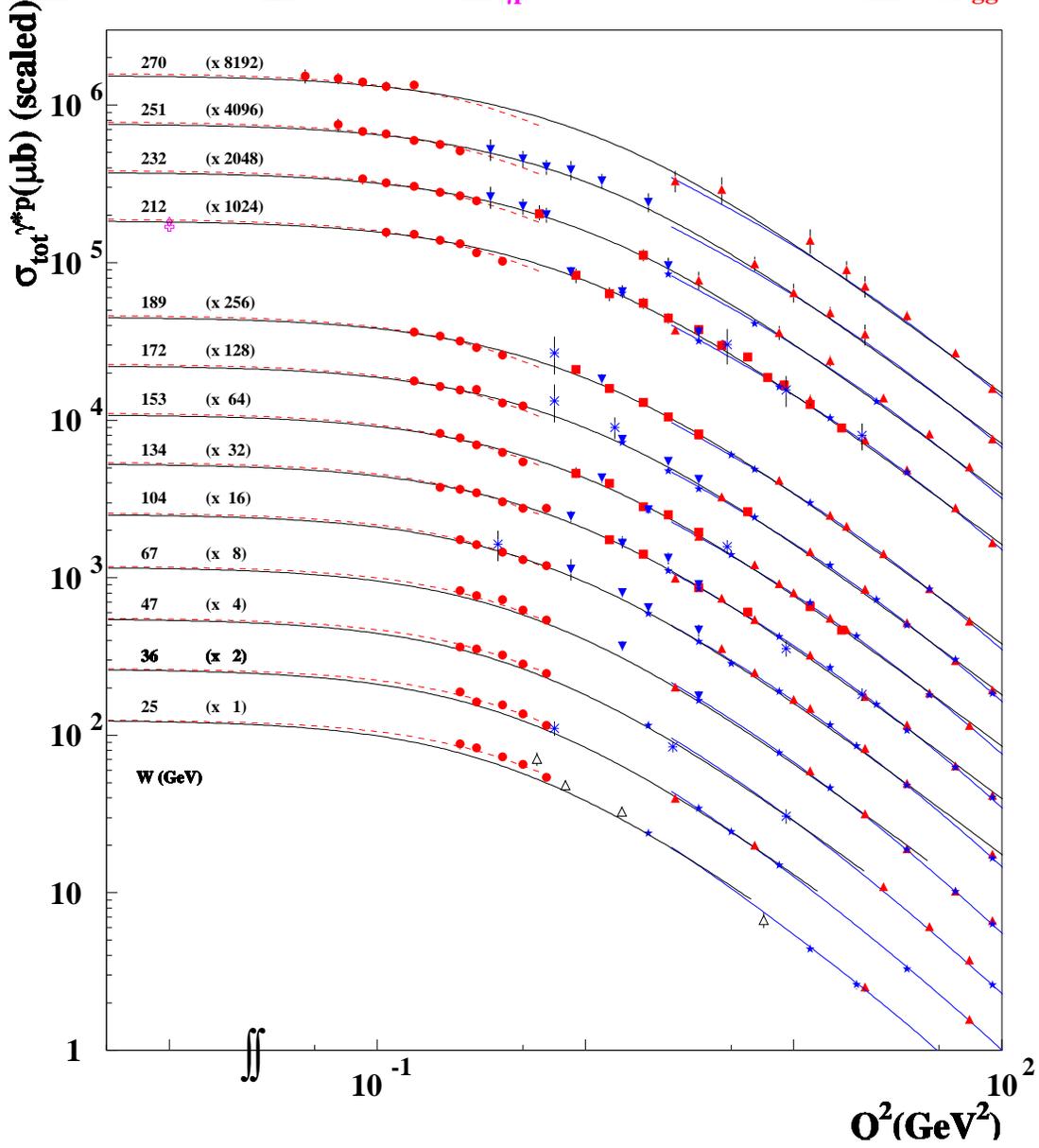,width=15.0cm}}
\caption{\it $\sigma_{tot}^{\gamma^{*} p}$ as a function of $Q^{2}$ for different values
of $W$.}
\label{sig_q2}
\end{figure}

\begin{figure}[ht]
\centerline{\epsfig{figure=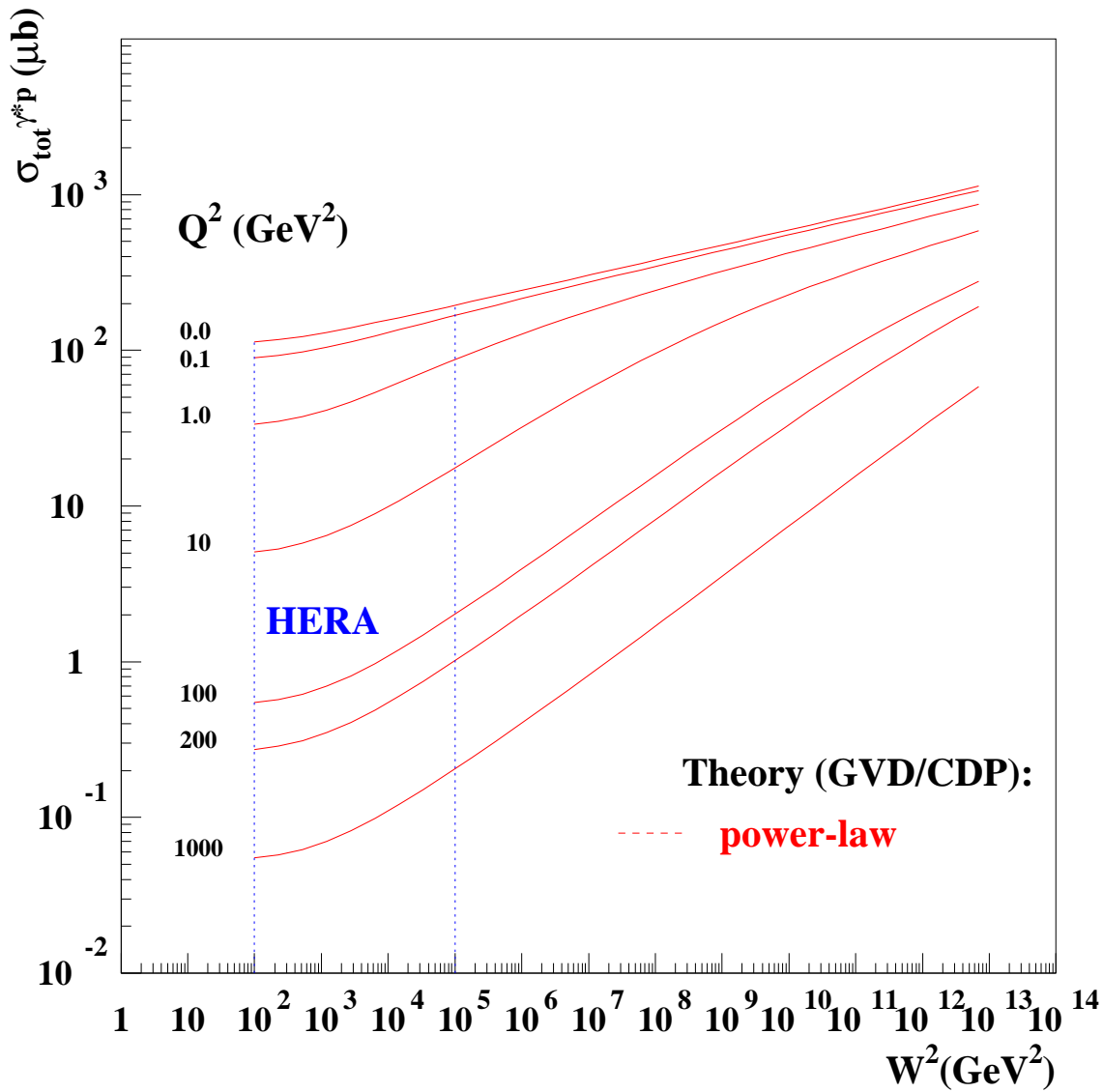,width=15.0cm}}
\caption{\it Expectations of the cross-section behavior of the 
total $\gamma^{*}p$ cross-section at large $W^{2}$ using the derived
analytical expression for $\sigma_{tot}^{\gamma^{(*)}p}$ by \cite{ref4}}
\label{sig_gvdm_w2}
\end{figure}

If one assumes that this behavior continues at extremely low $x$ or large values of $W^{2}$ 
one
is faced to conclude that $\sigma_{tot}^{\gamma^{*}p}$ reaches the energy behavior 
of the total photoproduction cross-section for any value of $Q^{2}$. 
Thus there are two limits which allows the 
total $\gamma^{*}p$ cross-section to saturate. First, for fixed values of $W^{2}$, one finds
$\sigma_{tot}^{\gamma^{*}p} \rightarrow \sigma_{tot}^{\gamma p}$ which has been clearly seen
at HERA (Figure \ref{sig_q2}). The second limit of $x \rightarrow 0$ or equivalently of
$W^{2} \rightarrow \infty$ at fixed $Q^{2}$ can be inferred from the above findings within the 
GVD/CDP picture.
The total $\gamma^{*}p$ cross-section reaches for $W^{2} \rightarrow \infty$ the
total photoproduction cross-section limit for any fixed values of $Q^{2}$. The second limit
however has not been experimentally confirmed so far and it will be rather difficult in the
future to do so. Figure \ref{sig_gvdm_w2} shows expectations for the behavior of
the total $\gamma^{*}p$ cross-section at large $W^{2}$ using the derived
analytical expression for $\sigma_{tot}^{\gamma^{*}p}$ by \cite{ref4}. 
This demonstrates clearly the asymptotic behavior 
$\sigma_{tot}^{\gamma^{*}p}/\sigma_{tot}^{\gamma p}\rightarrow 1$ for $W^{2}\rightarrow \infty$.
The kinematical region of HERA is also shown which is far from the second limit
described above. 

In summary, one is led to conclude based on the findings within the GVD/CDP picture
that the real and virtual photons on protons become identical in the limit of infinite energy.

\pagebreak


\end{document}